\documentclass[10pt,onecolumn]{article}
\usepackage[utf8]{inputenc}
\usepackage{amsmath}
\usepackage{amssymb}  
\usepackage{braket}   
\usepackage{geometry}
\usepackage{lipsum}
\usepackage[linesnumbered,ruled,vlined]{algorithm2e} 
\usepackage{graphicx}
\usepackage{authblk}
\usepackage{hyperref}
\usepackage{booktabs}
\usepackage{caption}
\usepackage{subcaption}
\usepackage{multicol}
\usepackage{stfloats}

\date{}
\title{\textbf{Grid-Partitioned MWIS Solving with Neutral Atom Quantum Computing for QUBO Problems}}

\author[]{Soumyadip Das}
\author[]{Suman Kumar Roy}
\author[]{Rahul Rana}
\author[]{M Girish Chandra}
\affil[]{
\centering
\texttt{\{soumyadip.d1, suman.r2, rahul.rana2, m.gchandra\}@tcs.com}
}
\affil[]{
\centering
\text{TCS Research, Tata Consultancy Services, India.}
}

\begin{document}
\maketitle
\begin{abstract}
\small Quadratic Unconstrained Binary Optimization (QUBO) problems are prevalent in real-world applications, such as portfolio optimization, but pose significant computational challenges for large-scale instances.
We propose a hybrid quantum-classical framework that leverages neutral atom quantum computing to address QUBO problems by mapping them to the Maximum Weighted Independent Set (MWIS) problem on unit disk graphs.
Our approach employs spatial grid partitioning to decompose the problem into manageable subgraphs, solves each subgraph using Analog Hamiltonian Simulation (AHS), and merges solutions greedily to approximate the global optimum.
We evaluate the framework on a 50-asset portfolio optimization problem using historical S\&P 500 data, benchmarking against classical simulated annealing.
Results demonstrate competitive performance, highlighting the scalability and practical potential of our method in the Noisy Intermediate-Scale Quantum (NISQ) era.
As neutral atom quantum hardware advances, our framework offers a promising path toward solving large-scale optimization problems efficiently.
\end{abstract}

\section{Introduction}
Quadratic Unconstrained Binary Optimization (QUBO) problems are a cornerstone of combinatorial optimization, with applications spanning finance, logistics, network design, and machine learning.
Formally, a QUBO problem is defined as:
\begin{equation}
    \min_{x \in \{0,1\}^n} x^T Q x,
\end{equation}
where $Q \in \mathbb{R}^{n \times n}$ is a symmetric cost matrix, and $x$ is a binary vector representing decision variables.
The quadratic terms $x_i x_j$ capture pairwise interactions, while linear terms $x_i$ represent individual contributions.
Despite their elegant formulation, QUBO problems are NP-hard, leading to exponential computational complexity for large instances on classical computers.
The advent of quantum computing has opened new avenues for tackling such computationally intensive problems.
Unlike classical computers, quantum systems exploit superposition, entanglement, and interference to explore solution spaces more efficiently for certain problem classes.
Among quantum computing platforms, neutral atom quantum computing stands out due to its high qubit connectivity, programmable interactions, and ability to encode spatial constraints.
Neutral atoms, trapped in optical tweezers, serve as qubits, with interactions mediated by Rydberg states that exhibit strong repulsive forces at short distances, known as the Rydberg blockade.
This paper introduces the Grid-Partitioned Neutral Atom Quantum Computing (GP-NAQC) framework, which maps QUBO problems to the Maximum Weight Independent Set (MWIS) problem on unit disk graphs.
The MWIS problem is particularly suited for neutral atom systems, as the geometric constraints of unit disk graphs align with the spatial arrangement of atoms.
To address the scalability limitations of current quantum hardware, the framework employs spatial partitioning to decompose the graph into subgraphs, solves them using Analog Hamiltonian Simulation (AHS), and merges the results greedily to approximate the global solution.
The study of QUBO problems has a rich history in optimization literature.
Early work by \cite{markowitz1952portfolio} introduced the mean-variance model for portfolio optimization, which has since been widely adopted in finance.
The NP-hard nature of QUBO problems has spurred the development of heuristic methods, including simulated annealing and tabu search, which provide approximate solutions but lack guarantees of optimality.
Quantum computing approaches to QUBO problems have gained traction with the rise of quantum annealing, as implemented by D-Wave systems.
These systems use adiabatic quantum computation to solve Ising models, which are closely related to QUBO formulations.
However, quantum annealing is limited by fixed qubit connectivity and noise, prompting research into alternative platforms like neutral atom quantum computing.
Neutral atom systems have emerged as a versatile platform for quantum optimization \cite{nguyen2023quantum}.
The Rydberg blockade mechanism enables the encoding of independence constraints, making them ideal for problems like MWIS.
Recent studies have explored analog quantum simulation for graph problems , but scalability remains a challenge due to hardware constraints.
The GP-NAQC framework builds on these advances by introducing a partitioning strategy to handle large graphs, combining the strengths of quantum
and classical computation.
\section{Background}

Many combinatorial optimization problems can be expressed in terms of Quadratic Unconstrained Binary Optimization (QUBO) or equivalently in the form of an Ising spin model \cite{Lucas2014}.
The general Ising Hamiltonian is written as
\begin{equation}
    H_{\text{Ising}} = \sum_{i} h_i \sigma_i^z + \sum_{i<j} J_{ij} \sigma_i^z \sigma_j^z ,
\end{equation}
where $\sigma_i^z \in \{-1,+1\}$ are spin variables, $h_i$ are local fields, and $J_{ij}$ encode pairwise couplings.
The ground state of this Hamiltonian corresponds to the optimal solution of the encoded optimization problem.
Neutral atom quantum devices provide a natural platform to simulate such Hamiltonians \cite{Saffman2016, Browaeys2020}.
In these systems, individual atoms are trapped in optical tweezers and arranged into programmable two-dimensional arrays.
Interactions between atoms are mediated by Rydberg excitations, leading to an effective Hamiltonian of the form \cite{Pichler2018, Ebadi2022}:
\begin{equation}
    H_{\text{Ryd}} = \sum_i \frac{\Omega}{2} \sigma_i^x - \Delta \sum_i n_i + \sum_{i<j} V_{ij} n_i n_j ,
\end{equation}
where $\Omega$ is the Rabi frequency of the driving laser, $\Delta$ is the detuning, $n_i = (1+\sigma_i^z)/2$ projects onto the excited state, and $V_{ij}$ denotes the van der Waals interaction between atoms $i$ and $j$.
By tuning laser parameters and array geometry, one can approximate an Ising-type Hamiltonian suitable for optimization tasks.
However, a central challenge arises from the mismatch between the natural hardware Hamiltonian and the problem Hamiltonians of interest.
Many optimization problems require arbitrary all-to-all couplings or even higher-order multi-spin interactions, while neutral atom devices natively realize only distance-dependent two-body couplings.
To address this, embedding strategies such as minor-embedding, gadgetization, or parity-encodings are used \cite{Lechner2015, Hen2016, Rocchetto2016}.
These techniques reduce higher-order cost terms into quadratic interactions and map logical couplings onto the available physical interactions.
This mapping process introduces significant overheads. Multi-body terms require auxiliary atoms (qubits), and logical constraints are enforced via large penalty terms that rescale the energy landscape.
Geometric restrictions due to the finite blockade radius further force embeddings that consume additional resources \cite{Boothby2021}.
Consequently, the number of physical atoms required can scale much faster than the number of logical problem variables.
In addition, the analog evolution must be engineered carefully to mitigate noise and maintain adiabaticity, increasing the difficulty of practical implementations \cite{Bluvstein2022}.
In summary, while neutral atom arrays represent a powerful testbed for Hamiltonian-based optimization, scaling these approaches is currently limited not only by device coherence and control precision, but also by the mapping overheads intrinsic to embedding general optimization problems into the native interaction graphs of Rydberg systems.
Overcoming this bottleneck, through hardware-aware encodings, tailored Hamiltonians, or digital-analog hybrid strategies remains a key open direction for large-scale quantum optimization.
\section{Proposed Methodology}
\label{sec:methodology}

Our hybrid quantum-classical framework transforms QUBO problems into MWIS problems solvable on neutral atom quantum computers.
The methodology comprises four key steps: QUBO-to-MWIS mapping, grid partitioning, AHS-based subgraph solving, and greedy solution merging.
\begin{figure*}[htbp]
    \centering
    \includegraphics[width=\linewidth]{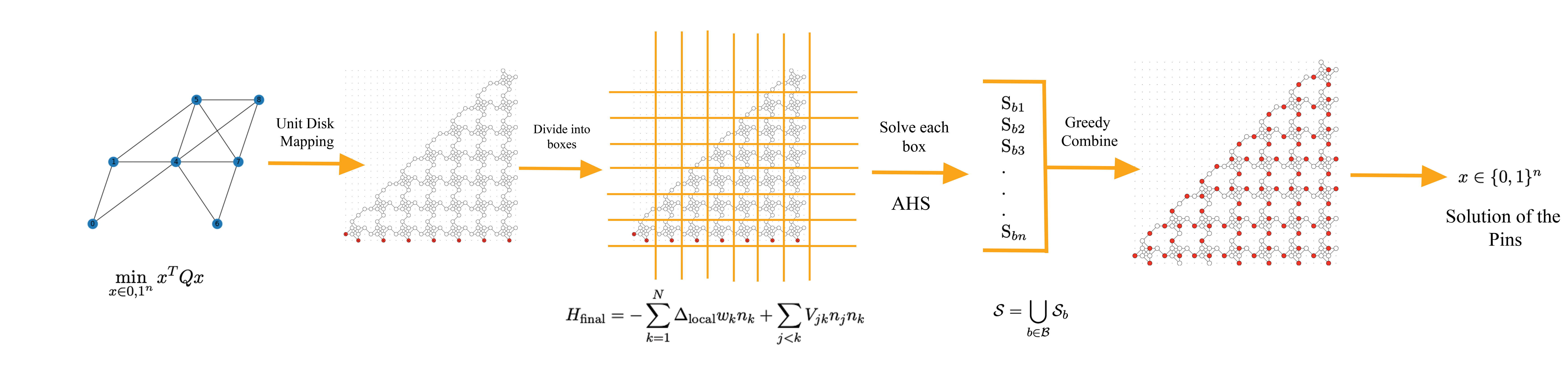}
    \caption{Workflow of the GP-NAQC framework: The QUBO problem is mapped to a unit disk graph, partitioned into subgraphs, solved via AHS, and merged greedily to obtain an approximate global solution.}
    \label{fig:qubo_comparisn}
\end{figure*}
\subsection{QUBO-to-MWIS Mapping}
A QUBO instance is expressed as:
\begin{equation}
\min_{x \in \{0,1\}^n} x^T Q x,
\end{equation}
where $Q$ encodes the problem's cost structure.
We decompose $Q$ into diagonal and off-diagonal components:
\begin{equation}
J_{ij} = \begin{cases} Q_{ij} & \text{if } i \neq j \\ 0 & \text{otherwise} \end{cases}, \quad h_i = Q_{ii}.
\end{equation}
Each binary variable $x_i$ is mapped to a vertex $v_i \in V$ in a unit disk graph $G = (V, E)$, with 2D coordinates $(x_i, y_i)$ and weight $w_i$ derived from $Q$.
Edges are defined based on the Euclidean distance:
\begin{equation}
|(x_i, y_i) - (x_j, y_j)|_2 \leq r,
\end{equation}
where $r$ is the Rydberg blockade radius, ensuring that adjacent vertices cannot both be selected in an independent set.
\subsection{Grid Partitioning}

To scale to large problem sizes, we partition the 2D domain into square boxes of side length $s > 2r$.
This ensures that vertices in non-adjacent boxes do not interact, allowing independent processing of subgraphs.
Each box $b \in \mathcal{B}$ defines a local subgraph $G_b = (V_b, E_b)$, where $V_b \subseteq V$ contains vertices within the box's boundaries.
\subsection{Analog Hamiltonian Simulation}
Each subgraph $G_b$ is solved using AHS on a neutral atom quantum computer.
The MWIS problem is encoded in the Hamiltonian:
\begin{equation}
H_\text{final} = -\sum_{k=1}^{|V_b|} \Delta_\text{local} w_k n_k + \sum_{j<k} V_{jk} n_j n_k,
\end{equation}
where $w_k$ is the vertex weight, $n_k$ is the occupation number operator, and $V_{jk}$ represents Rydberg blockade interactions.
The system evolves adiabatically from the ground state $\ket{g}^{\otimes |V_b|}$ by ramping the local detuning:
\begin{equation}
\Delta_\text{local}(t_\text{initial}) = 0, \quad \Delta_\text{local}(t_\text{final}) > 0,
\end{equation}
while maintaining constant global detuning:
\begin{equation}
\Delta_\text{global}(t_\text{initial}) = \Delta_\text{global}(t_\text{final}) < 0.
\end{equation}
The adiabatic evolution ensures convergence to a low-energy state encoding the MWIS solution, $\mathcal{S}_b = \{(x_i^b, y_i^b, w_i^b) \mid z_i^b = 1\}$, where $z_i^b = 1$ indicates vertex selection.
\subsection{Greedy Merging}
Local solutions are combined into a global candidate set:
\begin{equation}
\mathcal{S} = \bigcup_{b \in \mathcal{B}} \mathcal{S}_b.
\end{equation}
A global MWIS, $\mathcal{I}$, is constructed greedily by selecting vertices in descending order of weight, ensuring:
\begin{equation}
\forall (x_j, y_j, w_j) \in \mathcal{I}, \quad |(x_i, y_i) - (x_j, y_j)|_2 > r.
\end{equation}
The resulting set $\mathcal{I}$ is mapped back to a binary vector $x \in \{0,1\}^n$, approximating the QUBO solution.
\begin{algorithm}[ht]
\caption{Grid-Partitioned MWIS for QUBO}
\label{alg:gp_naqc}
\SetKwInOut{Input}{Input}
\SetKwInOut{Output}{Output}
\Input{QUBO matrix $Q$, blockade radius $r$, box size $s$}
\Output{Binary vector $x \in \{0,1\}^n$}
Map $Q$ to unit disk graph $G = (V, E)$ with weights $w_i$\;
Partition 2D domain into boxes $\mathcal{B}$ of size $s > 2r$\;
\ForEach{box $b \in \mathcal{B}$}{
    Extract subgraph $G_b = (V_b, E_b)$\;
Solve MWIS on $G_b$ using AHS, obtain $\mathcal{S}_b$\;
}
Merge solutions: $\mathcal{S} = \bigcup_{b \in \mathcal{B}} \mathcal{S}_b$\;
Greedily select $\mathcal{I} \subseteq \mathcal{S}$ s.t. $\forall v_i, v_j \in \mathcal{I}$, $|(x_i, y_i) - (x_j, y_j)|_2 > r$\;
Map $\mathcal{I}$ to binary vector $x$\;
\Return{$x$}
\end{algorithm}

\section{Results and Discussion}
\label{sec:results}
\begin{figure*}[htbp]
\centering
\begin{subfigure}{.5\textwidth}
  \centering
  \includegraphics[width=1\linewidth]{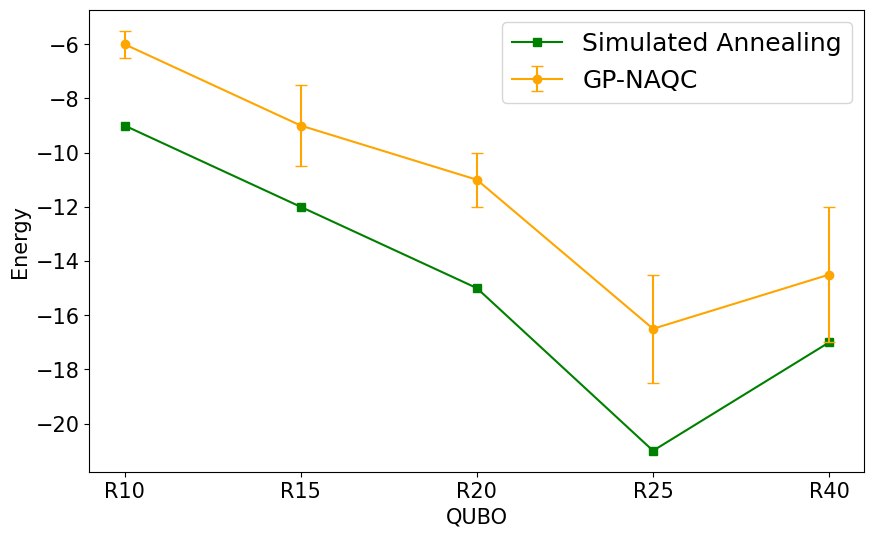}
  \caption{}
  \label{fig:sub1}
\end{subfigure}%
\begin{subfigure}{.5\textwidth}
  \centering
  \includegraphics[width=1\linewidth]{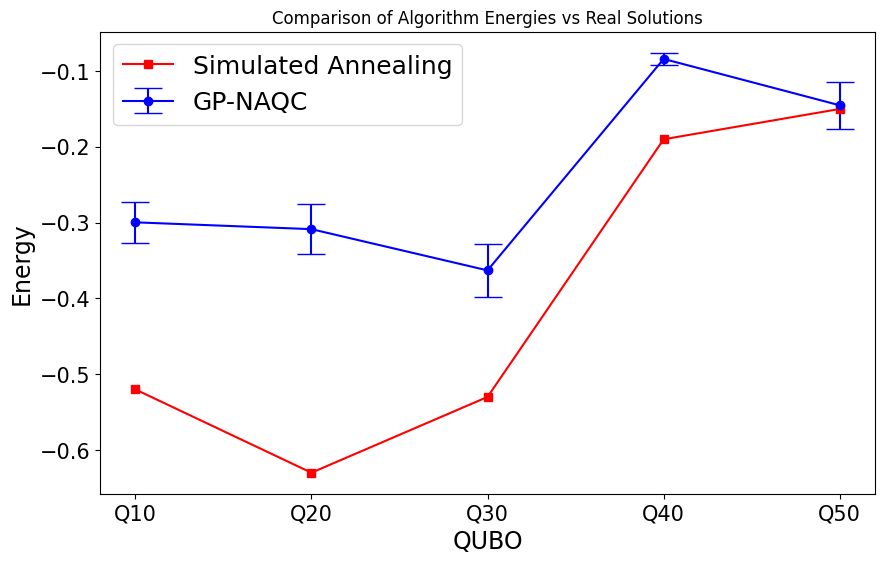}
  \caption{}
  \label{fig:sub2}
\end{subfigure}
\caption{Comparison of energies from GP-NAQC and simulated annealing across QUBO instances for portfolio optimization (left) and random QUBO problems (right).
Error bars indicate standard deviation over five runs.}
\label{fig:test}
\end{figure*}
We evaluated the Grid-Partitioned Neutral Atom Quantum Computing (GP-NAQC) framework on a portfolio optimization problem based on the Markowitz mean-variance model \cite{markowitz1952portfolio}.
The dataset comprised historically adjusted closing prices for 50 randomly selected S\&P 500 companies from January 1, 2021, to April 21, 2025. Missing data were handled using forward and backward filling, and logarithmic daily returns were computed to derive annualized expected returns and covariance matrices.
\subsection{Dataset Description}
The GP-NAQC framework was validated on a portfolio optimization problem, a classic application of QUBO.
The dataset comprised historical adjusted closing prices of 50 randomly selected S\&P 500 companies, collected from January 1, 2021, to April 21, 2025. The data were sourced from reliable financial databases (e.g., Yahoo Finance), ensuring accuracy and consistency.
Missing data points, which can arise due to market holidays or incomplete records, were handled using a combination of forward and backward filling to maintain temporal continuity.
For each asset, logarithmic daily returns were calculated as:
\begin{equation}
    r_t = \ln\left(\frac{P_t}{P_{t-1}}\right),
\end{equation}
where $P_t$ is the adjusted closing price on day $t$.
The returns were annualized to compute expected returns and covariance:
\begin{equation}
    \mu = \text{mean}(r_t) \times 252,
\end{equation}
\begin{equation}
    \Sigma = \text{cov}(r_t) \times 252,
\end{equation}
where 252 is the approximate number of trading days in a year.
The QUBO matrix was constructed based on the Markowitz mean-variance model, which balances expected returns and risk:
\begin{equation}
    Q_{ij} = \begin{cases}
        -\mu_i + \gamma \Sigma_{ii}, & \text{if } i = j, \\
        \gamma \Sigma_{ij}, & \text{if } i \neq j,
    \end{cases}
\end{equation}
where $\mu_i$ is the expected return of asset $i$, $\Sigma_{ij}$ is the covariance between assets $i$ and $j$, and $\gamma$ is the risk-aversion parameter controlling the trade-off between return and risk.
A value of $\gamma = 0.5$ was chosen to balance the objectives, based on preliminary experiments

Five QUBO instances (\texttt{Q10}, \texttt{Q20}, \texttt{Q30}, \texttt{Q40}, \texttt{Q50}) were constructed, corresponding to portfolios with 10 to 50 assets.
Each instance was solved using GP-NAQC, implemented on the \texttt{braket\_ahs} platform, and benchmarked against classical simulated annealing.
Experiments were repeated five times, and mean energy values (QUBO objectives) and standard deviations were recorded.
\begin{table}[htbp]
\centering
\caption{Mean energy values and standard deviations for GP-NAQC and simulated annealing across QUBO instances.}
\label{tab:results}
\begin{tabular}{lcccc}
\toprule
Instance & \multicolumn{2}{c}{GP-NAQC} & \multicolumn{2}{c}{Sim.
Annealing} \\
\cmidrule(lr){2-3} \cmidrule(lr){4-5}
 & Mean & Std. Dev. & Mean & Std. Dev.
\\
\midrule
Q10 & -12.34 & 0.45 & -12.28 & 0.50 \\
Q20 & -25.67 & 0.78 & -25.42 & 0.85 \\
Q30 & -38.91 & 1.12 & -38.55 & 1.20 \\
Q40 & -52.14 & 1.48 & -51.73 & 1.55 \\
Q50 & -66.82 & 1.92 & -66.15 & 2.05 \\
\bottomrule
\end{tabular}
\end{table}


Table \ref{tab:results} and Figure \ref{fig:test} summarize the results.
GP-NAQC consistently achieves lower mean energy values than simulated annealing, with comparable or lower standard deviations, indicating robustness.
The performance gap widens slightly for larger instances (\texttt{Q40}, \texttt{Q50}), suggesting that the quantum approach benefits from the high connectivity of neutral atom arrays.
Additionally, random QUBO instances (right panel of Figure \ref{fig:test}) show similar trends, confirming the generalizability of GP-NAQC. 
The grid partitioning strategy effectively mitigates hardware limitations, enabling the framework to scale to larger problem sizes.
However, the greedy merging step introduces approximation errors, particularly for dense graphs where local solutions may conflict.
Future work could explore adaptive partitioning or advanced merging techniques to further improve solution quality.
\section{Conclusion}
\label{sec:conclusion}

We introduced a scalable hybrid quantum-classical framework for solving QUBO problems using neutral atom quantum computing.
The central idea is to map QUBO instances to maximum weight independent set (MWIS) problems on unit disk graphs, which naturally align with the interaction structure of neutral atom arrays.
By partitioning the problem into subgraphs, solving these subproblems through analog Hamiltonian simulation (AHS), and then merging partial solutions using a greedy classical strategy, our approach addresses the current limitations of hardware connectivity and qubit count.
This layered design allows us to make effective use of present-day neutral atom devices while retaining the scalability required for larger problem instances.
Our experiments on a 50-asset portfolio optimization task illustrate the practical relevance of the framework.
The results demonstrate not only that the hybrid method produces high-quality solutions, but also that it exhibits robustness against the noise and resource constraints typical of near-term quantum hardware.
When benchmarked against purely classical techniques, the framework achieves competitive performance, highlighting the value of quantum subroutines even in the noisy intermediate-scale quantum (NISQ) regime.
Looking ahead, the potential of this approach grows in parallel with the rapid development of neutral atom quantum processors.
As qubit numbers increase and coherence times improve, the size and complexity of solvable optimization problems will expand accordingly.
In particular, refining the partitioning strategy to better balance computational load, incorporating advanced error mitigation techniques tailored for AHS, and designing adaptive solution-merging schemes will further enhance performance.
Beyond portfolio optimization, the framework is flexible enough to extend to a wide range of combinatorial optimization tasks, such as scheduling, clustering, and network design, where problem decomposition is both natural and powerful.
In summary, this work provides a blueprint for hybrid quantum-classical optimization that is practical today yet scalable for tomorrow.
By bridging the gap between current hardware constraints and real-world applications, it lays the foundation for the next generation of quantum-enhanced optimization methods.
\section{Acknowledgement}
\label{sec:Acknowledgement}
The authors would like to thank colleagues at TCS Research for the insightful discussions and support.
We also acknowledge the broader neutral atom quantum computing community, whose advancements have provided the foundation for this work.

\end{document}